\documentstyle[aps,prl,epsf,floats]{revtex}

\begin{document}

\ifpreprintsty \else
\twocolumn[\hsize\textwidth\columnwidth\hsize\csname@twocolumnfalse%
\endcsname \fi


\title{Multiple beam interference in a quadrupolar glass fiber}

\author{Martina Hentschel$^{(a)}$ and Matthias Vojta$^{(b)}$}
\address{
(a) Max-Planck-Institut f\"{u}r Physik komplexer Systeme,
N\"{o}thnitzer Str. 38, 01187 Dresden, Germany\\
(b) Theoretische Physik III, Elektronische Korrelationen und
Magnetismus, Universit\"at Augsburg, 86135 Augsburg, Germany
}

\date{June 20, 2001}
\maketitle

\begin{abstract}

Motivated by the recent observation of periodic filter
characteristics of an oval-shaped micro-cavity,
we study the possible interference of multiple beams
in the far field of a laser-illuminated quadrupolar glass
fiber.
From numerical ray-tracing simulations of the experimental
situation we obtain the interference-relevant
length-difference spectrum and compare it with
data extracted from the experimental filter
results.
Our analysis reveals that different polygonal cavity modes
being refractively output-coupled in the high-curvature
region of the fiber contribute to the observed
far-field interference.

\end{abstract}
\pacs{111.111}

\ifpreprintsty \else
] \fi              



{\em Introduction.}
Optical fibers and cavities
have attracted a lot of interest in recent years.
On the theoretical side, a plethora of phenomena
related to the interplay of classical and quantum
chaos is found~\cite{chaos,jens}.
On the experimental side, fibers are used
either as (active) lasing fibers in microlasers,
or as (passive) optical
filters which are of great technological interest
for planar integrated filter applications.
Planar dielectric ring and disc cavities have been used
as micron-sized optical filters mainly with
evanescent light coupling, working with nearly total
internal reflection.
However, evanescent coupling between the cavity curved
sidewall and the waveguide flat sidewall
requires a very precise fabrication
with a gap spacing in the sub-$\mu m$ range.
Therefore, filter techniques using non-evanescent coupling
which allows gap sizes larger than sub-$\mu m$ are
technologically desirable.

Interestingly, recent experiments \cite{poon} using an oval-shaped
micro-cavity have shown periodic output filter characteristics
which are potentially useful in the above context.
The purpose of this paper is to present an analysis
of the experimental data and numerical ray-tracing
simulations which allow a theoretical understanding
of these experimental findings.
Our results display a periodic filter spectrum in
the far-field interference of refractively output-coupled
modes of the cavity, in agreement with the experiment.
We will show that the interference of multiple
beam parts corresponding to polygonal round-trip orbits can
lead to the observed periodic filter characteristics in a narrow
window of the far-field angle.

First we briefly summarize the experiment of
Ref.~\onlinecite{poon}:
A passive (non-lasing) quadrupolar fiber
of high-refraction glass ($n=1.8$)
is illuminated by a laser beam, Fig.~\ref{fig:fiber}a.
The tunable laser source with wavelengths in the 670 $nm$ range
produces a Gaussian beam with
a width (spot size) of 30 $\mu m$, the
cavity axes are 150 and 180 $\mu m$.
The far-field elastic scattering spectrum is measured
with a linear array detector.
It
shows filter resonances as function of incoming
wavelength with a good peak to background ratio
of about 40, Fig.~\ref{fig:ftexp}a,
but only under very specific input and output
coupling angles $\theta_{i,o}$ (``magic window'').
The periodicity of the spectrum is a clear sign of
the interference nature of the phenomenon;
in addition
the filter peaks display inhomogeneous broadening
which is an indication of multimode interference.
The light wavelength $\lambda$ is much smaller
than the size of the cavity,
therefore quantum effects can be assumed to be small.

\begin{figure}
\epsfxsize=2.8in
\centerline{\epsffile{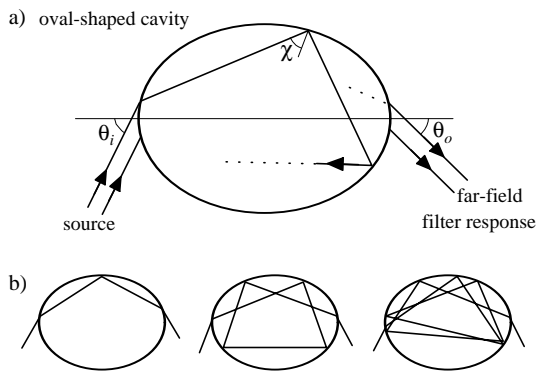}}
\caption{
a) Schematic experimental setup~\protect\cite{poon} for the
filter experiment. Shown is the cross section of the oval-shaped
fiber. If the input is coming from a broadband source, then the
output shows periodic filter characteristics, see also
Fig.~\protect\ref{fig:ftexp}. $\theta_{i,o}$ are the input and
output angles.
b) Typical orbits contributing to the far-field response, see text.
}
\label{fig:fiber}
\end{figure}


{\em Analysis of experimental data.}
If we accept that the periodic output characteristics
can be interpreted as interference of (classical) rays,
then a length analysis of the contributing geometric paths
appears suitable.
We start by defining the amplitude-weighted length
distribution $T(L)$ for the inferfering rays,
\begin{equation}
T(L) = \sum_{{\rm paths}\,i} A_i \, \delta(L - L_i)
\,,
\label{tl}
\end{equation}
where $A_i$ and $L_i$ are amplitude (at the detector) and
optical length (from source to detector) of each path $i$
hitting the detector.
The interference pattern $J(k)$ as function of the vacuum wavevector
$k = 2\pi/\lambda$ is given by
\begin{eqnarray}
  J(k) &=& \left| \int_{-\infty}^\infty {\rm d}L \, T(L) \, e^{i k L} \right|^2
        =:  \int {\rm d}l \, e^{-i k l}\, S(l)
\label{multbeamint} \:.
\end{eqnarray}
In the last step we have introduced the length-difference spectrum $S(l)$,
given by the self-convolution (correlation function) of $T(L)$,
$S(l) = \int {\rm d}L \, T(L) \,T(l+L)$;
for discrete paths with lengths $L_i$ the quantity $S(l)$ will be
non-zero for $l=L_i-L_j$ $\forall\, i,j$.
The length-difference spectrum $S(l)$ will be the main quantity of
our analysis, it is related by Fourier
transformation to the observed interference pattern $J(k)$;
the information about {\em absolute} path lengths does
of course not enter the interference result.

The Fourier transform $S(l)$ of a representative set of experimental
data from Ref.~\onlinecite{poon}
is shown in Fig.~\ref{fig:ftexp}.
We see large contributions to $S(l)$ at roughly equally
spaced $l$ values,
it is tempting to identify the spacing with the path length
of one round trip corresponding to a single dominating cavity orbit.
However, the intensity for larger $l$ is rather small, and the
broad peaks suggest that more than one orbit contributes to $S(l)$.
Similar results are obtained with other data sets of the
experiment~\cite{poon}.

At this point it is useful to discuss possible interference scenarios.
The simplest possibility is interference of rays
tunneling out of a single stable orbit that is traced over and
over again.
In this case $T(L)$ contains equally spaced peaks with
monotonically decreasing intensity ($\sim t (1-t)^{i-1}$ where
$t$ is the tunneling rate and $i$ the number of round trips);
a similar decay is then also
found in the peaks of the difference spectrum $S(l)$.
For the particular case of the quadrupolar fiber
a candidate stable orbit is the ``diamond'',
here the tunneling rate is rather small.
From this one would expect a slow, monotonous decay of the
peak intensities in $S(l)$ --
obviously this is not in agreement with the experimental data.
Furthermore, in the present experimental geometry
the diamond-like orbit cannot be excited by {\em refractive} input coupling,
but only by tunneling -- the resulting intensity is much lower than
for refractive coupling, and is too small
to account for the experimental observation.
Other orbits like rectangular/trapezoidal modes or
whispering-gallery (WG) orbits \cite{jens}
are either unstable or
cannot be excited with the present non-evanescent
coupling.
We are therefore lead to consider the interference of
rays from multiple orbits as explanation for the observations of
Ref.~\onlinecite{poon}.

\begin{figure}
\epsfxsize=2.8in
\centerline{\epsffile{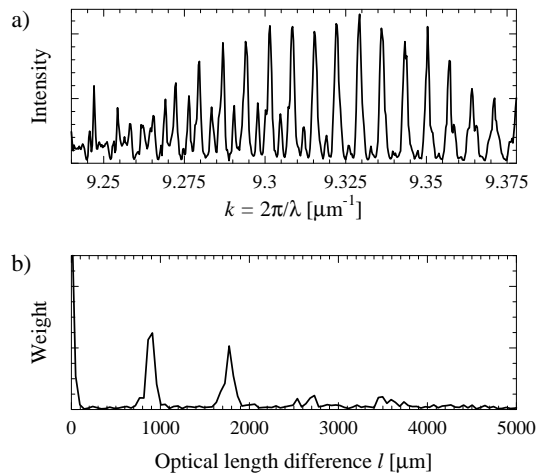}}
\caption{
Representative experimental spectrum of Ref.~\protect\onlinecite{poon}.
a) raw interference data, $J(k)$, plotted as function of the
wavevector $k$, $\theta_i=60^{\rm o}$, $\theta_o=56^{\rm o}$.
b) Modulus \protect\cite{realsym}
of the Fourier transform of $J(k)$ corresponding
to the length-difference spectrum $S(l)$.
The strong decay of the peak intensity for lengths $l$ longer
than 3 or 4 round trips is a characteristic feature.
}
\label{fig:ftexp}
\end{figure}



{\em Ray-tracing simulations.}
We focus a two-dimensional geometry representing the
cross-section of the fiber used in the experiment,
see Fig.~\ref{fig:fiber}a.
The geometry is defined by the mean radius $R$ of the cavity
($R=82\,\mu m$ in Ref.~\onlinecite{poon})
and its excentricity $\epsilon\approx 0.1$.
The quadrupole boundary is given by
$r(\phi) = R(1\!+\!\epsilon\cos 2\phi)$
in polar coordinates,
and the lengths of the half axes are
$R(1\pm\epsilon)$.

The incoming beam is discretized into a sufficiently high
number of equally spaced parallel rays, for simplicity we employ
a rectangular beam profile.
The intensity fraction of each ray that penetrates into the quadrupole
is given by the Fresnel formula, its angle by Snell's law.
Quantum effects are negligible due to $\lambda \ll R$.
The dynamics of each ray is then governed by the laws of a
``Fresnel billiard'' \cite{jens,berry},
{\em i.e.}, by straight propagation,
specular reflection at the quadrupolar shaped boundaries,
and evolution of the intensity according to Fresnel's law for
reflection and transmission.
In particular, for TE polarisation \cite{poon} the Fresnel laws
for the reflected (transmitted) electric field strength
$E_r$ ($E_t$) for a incoming field $E_i$ are
given by \cite{kapany}
\begin{eqnarray}
  q_r := \frac{E_r}{E_i} & = &
                \frac{ \cos\chi - n \sqrt{1 - n^2 \sin^2\chi} }
                { \cos\chi + n \sqrt{1 - n^2 \sin^2\chi} }
                \:, \label{refamp}\\
  q_t := \frac{E_t}{E_i} & = &
                 \sqrt{1 - q_r^2}
\,,
\label{trmamp}
\end{eqnarray}
where $n$ is the refractive index of the cavity material,
$n=1$ outside the cavity,
and $\chi$ is the boundary angle of incidence,
Fig.~\ref{fig:fiber}.
Strictly, these relations hold for a planar interface;
modified Fresnel formulae for curved interfaces can be
derived \cite{curved_fresnel}, but in the present case
of a very large size parameter, $R \gg \lambda$,
the curvature effect is negligible.
We assume perfectly reflecting walls for angles of
incidence $\chi$ bigger than the critical
angle, $\chi_c = \arcsin(1/n)$, and exclude leakage due to
quantum tunneling.

In the simulation, we trace each ray of the incoming beam
numerically to construct the interference pattern~\cite{jens,lock}.
For angles $\chi < \chi_c$ we allow for refractive escape
of the part of the ray that is
determined by Eq.~(\ref{trmamp}), 
but follow further the remaining part
inside the quadrupole until its intensity falls below a
threshold of $10^{-6}$ of the initial intensity due to
subsequent subcritical reflections.
For the transmitted part, we determine the far field angle
of the leaving ray again by Snell's law.

In Fig.~\ref{fig:fiber}b we show a couple of typical trajectories
that are found upon scanning of the incoming beam. 
Due to the finite excentricity and the finite beam width
we find not only WG orbits 
but a large number of orbits which escape (as well as enter)
around the points of highest curvature of the quadrupole,
as known from the study of asymmetric resonance cavities~\cite{jens,lock}.
In particular, we find rays that
undergo several polygonal-like round trips (in contrast to typical
WG orbits they come closer to the center of the quadrupole)
before their intensity eventually drops below the threshold.
This process of intensity loss may happen at one single
reflection or (more likely) upon a couple of subsequent
reflections/transmissions.
Due to the preferred escape points near the highest wall curvature~\cite{jens},
the distribution of the orbit lengths,
measured from entering the fiber until escape,
is sharply peaked at integer multiples of half the length of a
round trip.
Orbits with approximately 0.5, 1.5, 2.5 round trips are
shown in Fig.~\ref{fig:fiber}b -- these (and similar longer)
orbits contribute to the far-field interference for an output
angle as chosen in Fig.~\ref{fig:fiber}a, {\em i.e.}, for output
points opposite to the incident beam.

\begin{figure}[t]
\epsfxsize=2.7in
\centerline{\epsffile{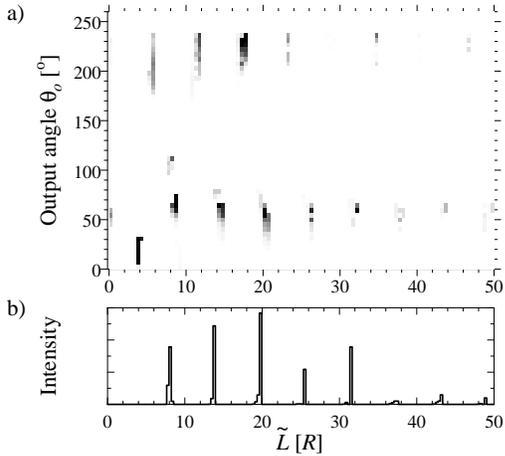}}
\caption{
Results of the ray-tracing simulation.
a) Intensity histogram showing the distribution of
orbit lengths $\widetilde L$ vs their output angle $\theta_o$.
The gray scale indicates the intensity (black - maximal).
The input angle is fixed at $\theta_i = 60^o$.
b) Intensity vs. orbit length $\widetilde{L}$ at a specific
detector position (in the ``magic window'', see text),
obtained by integrating the above histogram
over a narrow interval of output angles, $58^o < \theta_o < 62^o$.
If we neglect external path differences this quantity is
equivalent to the length distribution $T(L)$.
}
\label{fig:hist1}
\end{figure}

Given the experimental findings, we are especially interested in
rays with escape direction within the ``magic window''.
The filter characteristics was observed in the far-field
total intensity -- in this situation
the detector is placed at a distance large compared to the cavity
radius.
Therefore only rays within a narrow window of the output
{\em angle} $\theta_o$ are detected, whereas the exact escape
{\em position} on the quadrupolar boundary is of no importance.
In contrast, in a near field measurement (usually done with a focussing
lens) a narrow interval of output positions is sampled, with a
rather large range of output angles.

We now turn to the ray-tracing simulation results for the
far-field interference.
The primary output of the simulation are the ray
trajectories, their escape points and output angles,
and the output (transmission) intensities which result from
the Fresnel formulae (\ref{trmamp}).
Fig.~\ref{fig:hist1}a shows lengths $\widetilde{L}$ and output angles $\theta_o$
in the form of a intensity histogram plot (``Fresnel-weighted'' histogram)
for a fixed input angle $\theta_i = 60^o$.
Here, $\widetilde{L}$ are the geometric orbit lengths inside
the cavity in units of $R$; the optical lengths
are found from $L = n \widetilde{L} R + L_{\rm ext} + L_{\rm phase}$,
where $L_{\rm ext}$ are external length differences arising
from the different input and output points of the interfering rays,
and $L_{\rm phase}$ is determined by the phase shifts
that occur upon the 
reflections.
Note that the quantity shown in Fig.~\ref{fig:hist1}a
is the output-angle-resolved version of the
length spectrum $T(L)$ defined in (\ref{tl}).
For most output angles, the intensities are rather small, and
arise primarily from short lengths $\widetilde{L}$.
However, in a narrow window which corresponds to escape points
in the high-curvature region, the total output intensity is larger,
and short as well as longer orbits carry significant weight.

\begin{figure}
\epsfxsize=2.8in
\centerline{\epsffile{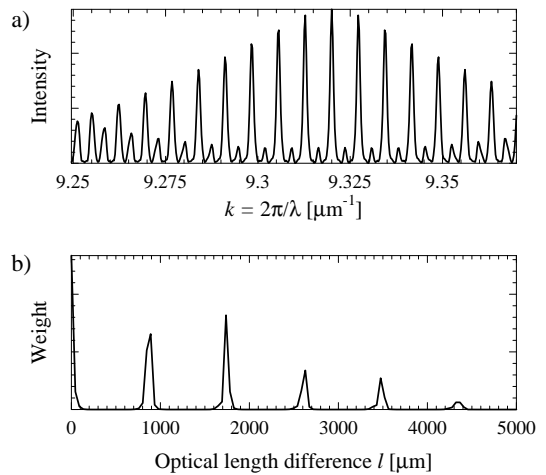}}
\caption{
Theoretical ray-tracing result, for the parameters $\theta_i = 60^o$,
$\theta_o = 60^o$, and $\epsilon =$ 0.1,
for details see text.
The axes are as in Fig.~\protect\ref{fig:ftexp}.
}
\label{fig:fttheor}
\end{figure}

If we now integrate the above quantity over a small range
of far-field angles corresponding to the angle range convered
by the detector, we obtain the length spectrum $T(\widetilde{L})$,
shown in Fig.~\ref{fig:hist1}b.
The peaks are easily found to correspond to orbits as shown
in Fig.~\ref{fig:fiber}b (and longer ones).
After converting $\widetilde L$ into optical lengths $L$,
we obtain from $T(L)$ the length-difference spectrum $S(l)$
by self-convolution, and by Fourier transformation \cite{FFT}
the resulting interference pattern
$J(k)$ as function of the wavevector $k=2\pi/\lambda$,
Fig.~\ref{fig:fttheor}.

Qualitative agreement between experiment and theory
is most easily checked
by comparing the length-difference spectra $S(l)$,
Figs.~\ref{fig:ftexp}b and \ref{fig:fttheor}b.
The main feature, namely a number of roughly
equally spaced peaks with comparable intensity,
but very little intensity at larger lengths,
is nicely reproduced by the ray-tracing data.

We note that one cannot expect to reach a quantitative agreement
with experiment due to the uncertainty in experimental
parameters like size and excentricity of the fiber and
angle of the illuminating laser beam.
The precise interference pattern depends strongly on
these parameters, in particular it is extremely
sensitive to length changes of the order of the light
wavelength.
For this reason we have neglected both $L_{\rm ext}$ and
$L_{\rm phase}$ when plotting Fig.~\ref{fig:fttheor}.
These contributions to the optical length are of the
order of one or several wavelengths, and are certainly
smaller than the uncertainty in the fiber size.
We have checked that the interference pattern depends
only slightly on the numerical discretization procedure
used for the incoming beam.

By varying input and output angles,
our simulation data clearly show
that in most situations the rays
from all parts of the incident beam travel in polygonal
cavity orbits only for a very short time (up to 2 round trips)
before they leave the cavity via refraction.
Only for a narrow range of input angles an appreciable
part of the beam leads to polygonal orbits with
a longer lifetime, which then are refractively
output-coupled after a larger number of round trips
into a narrow window of far-field angles.
The far-field output angle window depends sensitively
on the input angle, so we predict that the ``magic window''
as observed in Ref.~\onlinecite{poon} will move (in the
far-field angle) with varying input angle of the beam.
In particular, for $\theta_i=70^o$ we found the magic
window at output angles $10^o$ smaller than for $\theta_i=60^o$,
whereas for $\theta_i=50^o$ the filter effect almost
disappears.

In addition, a number of conditions have to be met
to observe the cavity filter effect:
(i) A finite excentricity of the fiber is
needed to produce (chaotic) orbits which come close
to the center of the quadrupole and leave it by refraction
preferrably near the points of highest curvature.
(ii) The intensity loss per round trip should be
neither too small nor too large.
In the former case, too many individual orbits
(with slightly different lengths) contribute
to the far-field interference leading to an incoherent
response, whereas in the latter case
the number of contributing beams is too small
to produce a sharp interference pattern.
This puts constraints on the refractive index of
the fiber.
We have performed simulations with other fiber
geometries and refraction indices which confirm the
points (i) and (ii) above: using {\em e.g.} $n=1.5$
the filter effect disappears.

{\em Summary.}
Our ray-tracing model appears to describe
the main features found in the experiment \cite{poon}.
In particular, the range of input and output angles, where
far-field interference with filter characteristics
can be observed, is rather small (``magic window'').
The analysis of the length-difference spectrum allows for
a clear distinction between our model of
interfering rays from different orbits and
other scenarios involving a single orbit only.
We note that the TE polarization used in Ref.~\onlinecite{poon}
allows for a very efficient Brewster-angle input and output
fiber coupling.

Issues to be discussed in future work include
(i) a comparison to the case of TM polarization,
(ii) the precise connection of the observed interference
to chaotic cacity trajectories, and
(iii) possible application of the filter effect
for beam and/or cavity diagnostics.

We are grateful to A.~W.~Poon for providing the
experimental data of Ref.~\onlinecite{poon},
furthermore we acknowledge useful discussions with
R.~K.~Chang, J.~U.~N\"ockel, A.~W.~Poon, A.~D.~Stone,
and J.~P.~Wolf.
This work has been supported by the DAAD and
the DFG.



\begin{references}
\bibitem{realsym} By definition, $J(k)$ is real and symmetric,
$J(k)=J(-k)$, and the same holds for $S(l)$, since $T(L)$ is real.
However, the numerical Fourier transform of $J(k)$, taken from
a finite interval $0<k_1<k<k_2$, can acquire a phase factor,
which is removed by working with the modulus.
\end{references}
\end{document}